\documentclass[prc,twoside,twocolumn]{revtex4}
\usepackage{srcltx}
\usepackage{graphicx}
\usepackage{epsfig,bm,dcolumn}
\usepackage{epstopdf}
\usepackage[utf8]{inputenc}
\usepackage{xcolor}

\newcommand{\ea}{\textit{et al.}}

\begin{document}

%%%%%%%%%%%%%%%%%%%%% Title %%%%%%%%%%%%%%%%%%%%%%
\title{Pion and Proton Form Factors in the Regge Description of Electroproduction $p(e,e'\pi^+)n$}

%%%%%%%%%%%%%%%%%%%% Authors %%%%%%%%%%%%%%%%%%%%%
%%%%%%%%%%%%%%%%%%%% Addresses %%%%%%%%%%%%%%%%%%%%%
\author{Tae Keun Choi}%
\email{tkchoi@yonsei.ac.kr} \affiliation{Department of Physics,
Yonsei University, Wonju 26493, Korea}
%--------------------------------------------------
\author{Kook Jin Kong}
\email[E-mail: ]{kong@kau.ac.kr} \affiliation{Research Institute
of Basic Sciences,
    Korea Aerospace University, Goyang 10540, Korea}
%--------------------------------------------------
\author{Byung Geel Yu}
\email[E-mail: ]{bgyu@kau.ac.kr} \affiliation{Research Institute
of Basic Sciences,
    Korea Aerospace University, Goyang 10540, Korea}
%--------------------------------------------------
%\date{\today}
%\date{}

%%%%%%%%%%%%%%%%%%%% Abstract %%%%%%%%%%%%%%%%%%%%%

\begin{abstract}
Electroproduction of $\pi^+$ above the resonance region is analyzed in
the Regge model for $\pi+\rho$ exchanges. The importance of the roles
of the pion and the proton form factors in the process is discussed in
comparison with the existing models of Kaskulov and Mosel and of Vrancx
and Ryckebusch. The present model with a proton
form factor of a simple dipole-type is shown to yield a better description
of DESY and JLab data over those models for the high $Q^2$ and
$-t$ region up to 5 GeV$^2$.
\end{abstract}

\maketitle

%\section{Introduction}

Recent experiments on the electroproduction of $\pi^+$ at JLab
has drawn attention because the data from the experiment cover a wide
range of the photon momentum transfer $Q^2$ at high energy $W$ 
\cite{Blok:2008jy,Horn:2007ug,park}. Analyses of electroproduction
data have shown that the cross sections of the process are largely
determined by the pion and the nucleon electromagnetic form factors
that include information about the hadron structure
\cite{vgl,kaskulov,vrancx}. Therefore, electroproduction plays a
role not only in understanding the production mechanism  but also in
testing various sorts of form factors originating either from
phenomenological or from some theoretical basis on Quantum Chromodynamics (QCD)
\cite{kaskulov,gari1}.

One interesting feature of the process is that empirical data at
large momentum transfer $Q^2$ and high energy $W$ require a large
contribution of the $s$-channel proton pole to the transverse
cross section $d\sigma_T/dt$ in order to reproduce a set of four
separated differential cross sections,
 %\begin{widetext}
    \begin{eqnarray}\label{crss1}
    &&2\pi\frac{d \sigma}{dt d\phi}= \frac{d\sigma_T}{dt}
    +\epsilon \frac{d\sigma_L}{dt} +\epsilon
    \frac{d\sigma_{TT}}{dt} \cos{2\phi}
    \nonumber\\&&
    %\hspace{1.5cm}
    +\sqrt{2\epsilon(\epsilon+1)}\frac{d\sigma_{LT}}{dt}
    \cos{\phi}\,,
    %&&\frac{d \sigma_{i}}{dt}=(\frac{\pi}{qk})\frac{d \sigma_i}{d\Omega_{\pi}} ,\quad
    %i=T,L,TT,TL \nonumber
    \end{eqnarray}
 %\end{widetext}
in a consistent manner. The longitudinal cross section
$d\sigma_L/dt$ follows the pion-pole dominance with the form
factor $F_\pi(Q^2)$ of a monopole type  from $\rho$-meson
dominance, as expected from
\begin{eqnarray}
d\sigma_L\propto \left|g_{\pi NN}F_\pi(Q^2)\right|^2.
\end{eqnarray}
For an enhancement of  $s$-channel exchange, therefore, in the
Regge model for $t$-channel meson exchange, Kaskulov and Mosel
(KM) \cite{kaskulov} introduced an $s$-dependence of the charge form
factor $F_s$ to the proton pole term to implement the contributions of
the $N^*$ resonances based on the Bloom-Gilman duality \cite{bloom},
i.e.,
\begin{eqnarray}\label{kmff}
F_s(Q^2,s)={\int_{M_p^2}^\infty ds_i{s_i^{-\beta}\over
s-s_i+i0^+}\left(1+\xi{Q^2\over s_i}\right)^{-2}\over
\int_{M_p^2}^\infty ds_i{s_i^{-\beta}\over s-s_i+i0^+}}\ ,
\end{eqnarray}
%The  choice of $\beta=3$ and $\xi =0.4$ fitted to the experiment
%dat from JLAB, DESY and Cornell to deep inelastic region at
%HERMES, and then the inegration of Eq. (\ref{kmff}) becomes
which results in Eq. (43) of Ref. 5. %\cite{kaskulov}.
%The corresponding form factor $F_p(Q^2,s)$ reads
%\begin{eqnarray}
%{s{\rm ln}\left[{\xi Q^2\over M_p^2}+1\right]{2\xi Q^2+s\over
%\xi^2 Q^4}-{s(\xi Q^2+s)\over \xi Q^2(\xi Q^2+M_p^2)}+{\rm
%ln}\left[s-M_p^2\over M_p^2\right]-i\pi}\over{\left({\xi Q^2\over
%s}+1\right)^2\left({s^2+2sM_p^2\over 2M^4_p}+{\rm
%ln}\left[{s-M_p^2\over M_p^2}\right]-i\pi\right)}.
%\end{eqnarray}
Within a framework similar to that of KM, Vrancx and Ryckebusch (VR)
showed a fit of cross sections by using the proton charge form factor
$F_s$ of a dipole type  \cite{vrancx},
\begin{eqnarray}\label{vrff}
F_p(Q^2,s) = \left(1 + Q^2/\Lambda^2_{\gamma pp^{*}}(s)\right)^{-2},
\end{eqnarray}
in which case the cutoff mass has an energy $s$-dependence.
As shown in Ref.~6, however, the validity of such form
factors of proton in these works are questionable for the large $-t$
region, which is particularly true for  very recent data \cite{park}.
\\

In this work we investigate the process
\begin{eqnarray}\label{process}
\gamma^*(k)+ p(p)\to\pi^+(q)+ n(p')
\end{eqnarray}
based on an extension of our previous works \cite{bgyu1} to
electroproduction. Our purpose here is to find another set of
possible parameters for the pion and the proton charge form factors to fit
the cross sections. Avoiding complications in the parameterization
of the proton charge form factor as in Eqs. (\ref{kmff}) and
(\ref{vrff}), we reconsider the proton form factor of a simple
dipole type as in Eq. (\ref{vrff}), but now it has a constant cutoff
mass  that can be adjusted.  We then examine whether that proton form factor is valid up to
$-t\approx 5$ GeV$^2$ at large $Q^2$ and high energy $W$.

\begin{figure}[]
    \centering
    \includegraphics[width=0.9\columnwidth]{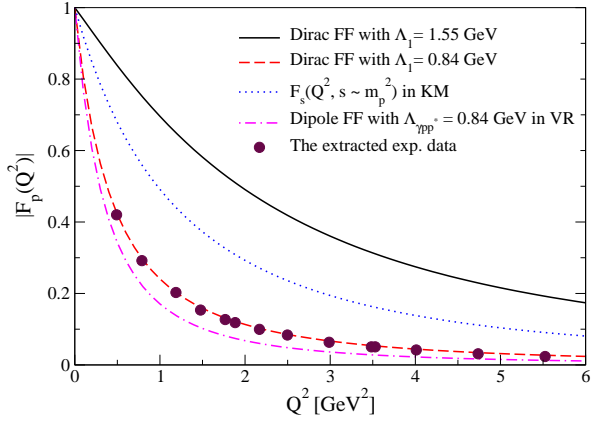}
    \caption{(Color online) The $Q^2$-dependence of
    proton charge form factors used in the models of Refs. 5 and 6 %\cite{kaskulov,vrancx}
    and the present work. The curves from KM and VR are estimated at
    $s=M_p^2$.
   }
    \label{emffs}
\end{figure}

For clarity, we work with a simple model consisting of $\pi+\rho$
Regge pole exchanges. In order to maintain gauge invariance of the
production amplitude, we have used the Gross-Riska prescription \cite{gross} 
for proton and  pion charge form factors to constrain
the respective longitudinal components from coupling to virtual photon.
Thus, the production amplitude can be written as
\begin{eqnarray}\label{born}
\mathcal{M}=\bar{u}_N(p')\sqrt{2}\left[{\cal M}_{s,p} +{\cal
M}_{t,\pi} +{\cal M}_{t,\rho} \right] u_N(p),
\end{eqnarray}
with
\begin{eqnarray}\label{s}
    &&i{\mathcal M}_{s,p}=e g_{\pi NN}
    \gamma_5 {(\rlap{/}{p}+\rlap{/}{k}+M_p)\over s-M_p^2}
     \widetilde{F}_{1}(Q^2)\rlap{/}\epsilon\,,\\
    &&i{\mathcal M}_{t,\pi}=e g_{\pi NN}
    \widetilde{F}_{\pi}(Q^2){(2q-k)\cdot\epsilon\over t-m^2_\pi}\,\gamma_5 ,\\
    &&i{\mathcal M}_{t,\rho}=-i g_{\gamma\rho\pi} g_{\rho NN}
    F_{\rho}(Q^2)\epsilon^{\mu\nu\alpha\beta}{\epsilon_\mu k_\nu q'_\alpha \over
    t-m^2_\rho} \nonumber\\&&\hspace{1cm}\times
    \left( \gamma_\beta+\frac{\kappa_\rho}{4m_p}[\gamma_\beta,\rlap{/}{q'}]
    \right),
\end{eqnarray}
where $\widetilde{F}_{1}(Q^2)\rlap{/}{\epsilon}$ and
$\widetilde{F}_{\pi}(Q^2)(2q-k)^\mu$ are
\begin{eqnarray}
&&(F_1(k^2)-F_1(0))
\left(\rlap{/}\epsilon-\rlap{/}{k}{k\cdot\epsilon\over k^2}\right)
+ F_1(0)\rlap{/}\epsilon,\label{gr-1}
\\
&&(F_{\pi}(k^2)-F_{\pi}(0)) (2q-k)\cdot\left(\epsilon-
k{k\cdot\epsilon\over k^2}\right)
\nonumber\\&&\hspace{1cm}
+F_{\pi}(0)(2q-k)\cdot\epsilon,
\label{gr-2}
\end{eqnarray}
respectively. $Q^2=-k^2$ is the virtual photon momentum and $q'=q-k$
is the $t$-channel momentum transfer.  The gauge-invariant $\rho$
meson exchange is denoted by ${\cal M}_{t,\rho}$, with the
transition form factor $F_\rho(Q^2)$ at the $\gamma\rho\pi$
coupling vertex with the coupling constant
$g_{\gamma\rho\pi}=0.22$ GeV$^{-1}$. For the meson-baryon coupling
constants, we use $g_{\pi NN}=13.4$ and $g_{\rho NN}=3.4$ with
$\kappa_\rho=6.1$.
\\
\begin{figure}[]
    \centering
    \includegraphics[width=0.9\columnwidth]{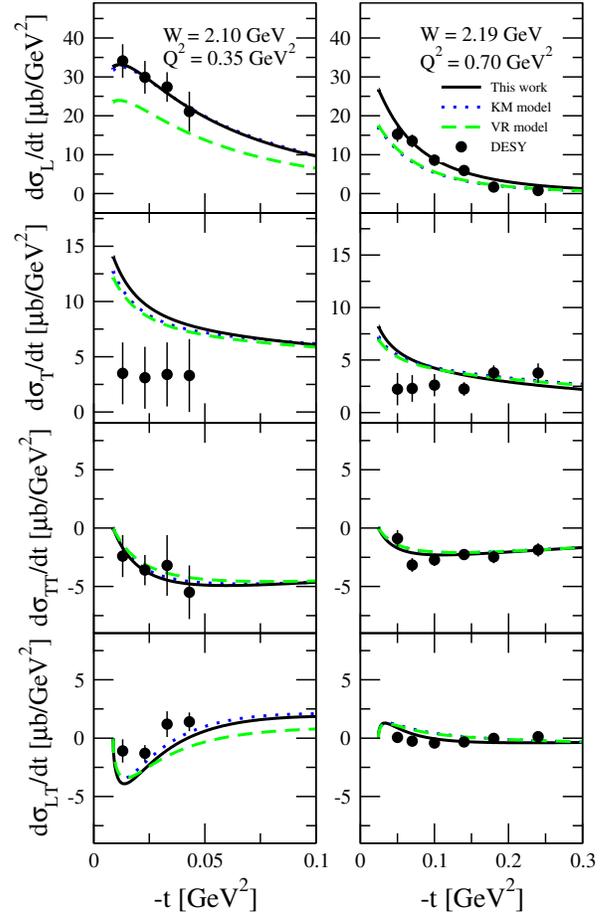}
    \caption{(Color online) Differential cross sections at very forward
    angles $-t<0.3$ GeV$^2$ and small $Q^2$.
    The solid line results from the present work with $\Lambda_\pi=0.78$ GeV and
    $\Lambda_1=1.55$ GeV.
    The dotted line is from the KM model with $\Lambda_\pi=0.775$ (left) and 0.63 (right) GeV
    and $F_s(Q^{2},s)$ in Eq. (43)
    in Ref. 5. %\cite{kaskulov}. 
    The dashed line is from the VR model with $\Lambda_\pi=0.655$ GeV and
    $F_p(Q^2,s)$ in Eq. (\ref{vrff}).
   Data are taken from Ref. 14 %\cite{ackermann}
   for $Q^2=0.35$ GeV$^2$ and from Ref. 15 %~\cite{brauel} 
    for $Q^2=0.7$~GeV$^2$ (DESY).
%\vspace{-0.5cm}
} \label{desy21}
\end{figure}

To reggeize the $t$-channel meson exchange, we simply replace the
Feynmann pole with the Regge pole which can be collectively written
as
\begin{eqnarray}
%\frac{1}{t-m_\varphi^2}\to
%&&
&&{\cal R}^\varphi(s,t)\nonumber\\
&&=\frac{\pi\alpha'_\varphi}{\Gamma(\alpha_\varphi(t)+1-J)}
\frac{{1\over2}((-1)^J+e^{-i\pi\alpha_\varphi})}{\sin\pi\alpha_\varphi(t)}
\left(\frac{s}{s_0}\right)^{\alpha_\varphi(t)-J}
\end{eqnarray}
for the $\varphi$ meson of arbitrary spin $J$. In the present work,
we use the complex phase $e^{-i\pi\alpha(t)}$ for both the $\pi$ and
the $\rho$ exchanges with the trajectories
\begin{eqnarray}
\alpha_\pi (t) &=& 0.7(t-m_\pi^2)\,, \\
\alpha_\rho (t) &=& 0.83t+0.53\,,
\end{eqnarray}
respectively.
\\

\begin{figure*}[]
    \centering
    \includegraphics[width=1.33\columnwidth]{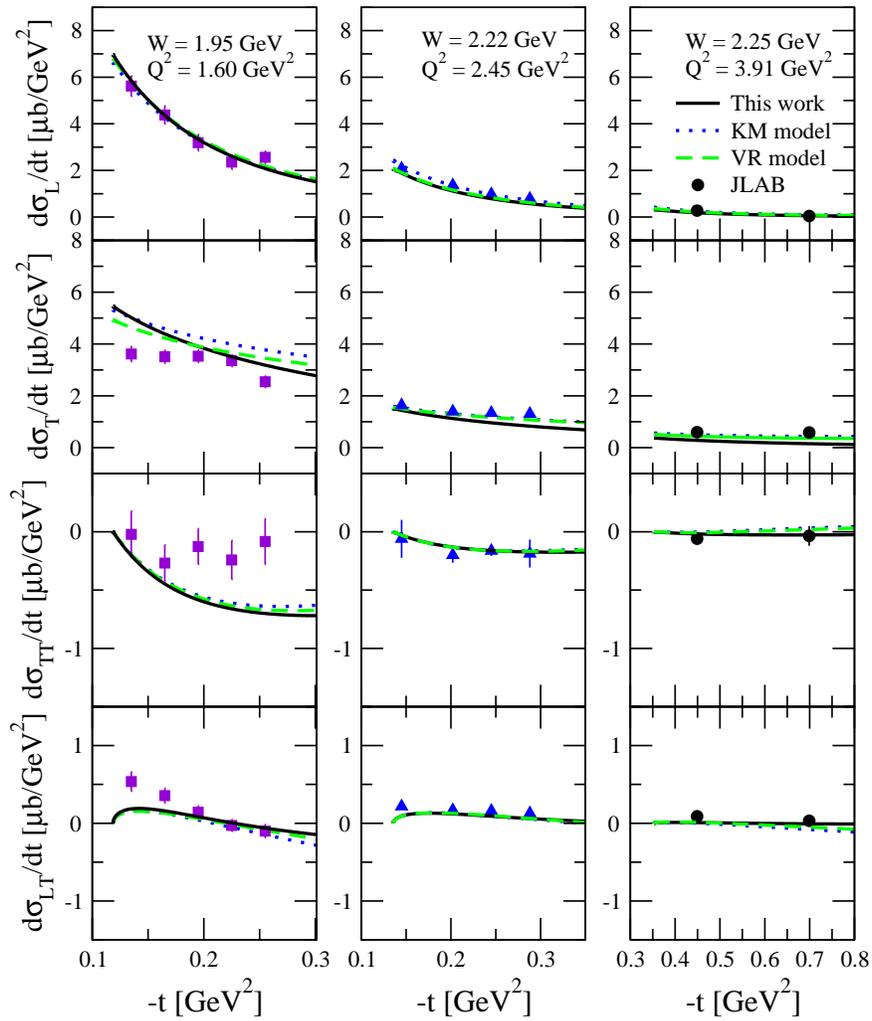}
    \caption{(Color online) Differential cross sections at forward angles $-t<1$ and
    $Q^2$ up to 4 GeV$^2$.
 The solid line results from the present work with $\Lambda_\pi=0.65$ GeV and $\Lambda=1.55$ GeV.
   Notations are the same as in Fig. \ref{desy21} for the dotted line with $\Lambda_\pi=0.68$ GeV
   and the dashed line with $\Lambda_\pi=0.655$ GeV.
   Data are taken from
   the F$\pi$-1 \cite{tadevosyan} (squares),
   F$\pi$-2 \cite{horn,Blok:2008jy} (triangles), and
   $\pi$-CT \cite{Horn:2007ug} (circles) experiments at JLab.}\label{jlab3exp}
\end{figure*}

Let us now discuss how to choose the pion and the proton charge form
factors in Eqs. (\ref{gr-1}) and (\ref{gr-2}), which are
needed to model the $\pi^+$ electroproduction process. In
the small $-t$ and low $Q^2$ region (see Fig. \ref{desy21} below,
for instance), cross section data show the longitudinal cross
section $d\sigma_L/dt$ to be large in comparison to the others. This is a
manifestation of the dominance of the pion exchange with the
charge form factor
\begin{eqnarray}\label{piff0}
F_{\pi}(Q^2)=\frac{1}{1+Q^2/\Lambda_\pi^2}\ ,
\end{eqnarray}
which is parameterized as a monopole-type from the vector meson
dominance. The cutoff mass is, therefore,  $\Lambda_\pi=m_\rho$,
which is somewhat larger than $\Lambda_\pi=0.71$ GeV fitted to the
measurement of the on-shell form factor in the elastic $e$N scattering
process. In the $\gamma^*p\to \pi^+n$ process, however, the pion
exchange proceeds via off-shell propagation; we, thus, consider
$\Lambda_\pi$ to be a fitting parameter to be varied in the range 0.6$\sim$
0.8 GeV  in this work.

\begin{figure*}[]
\centering
    \includegraphics[width=1.33\columnwidth]{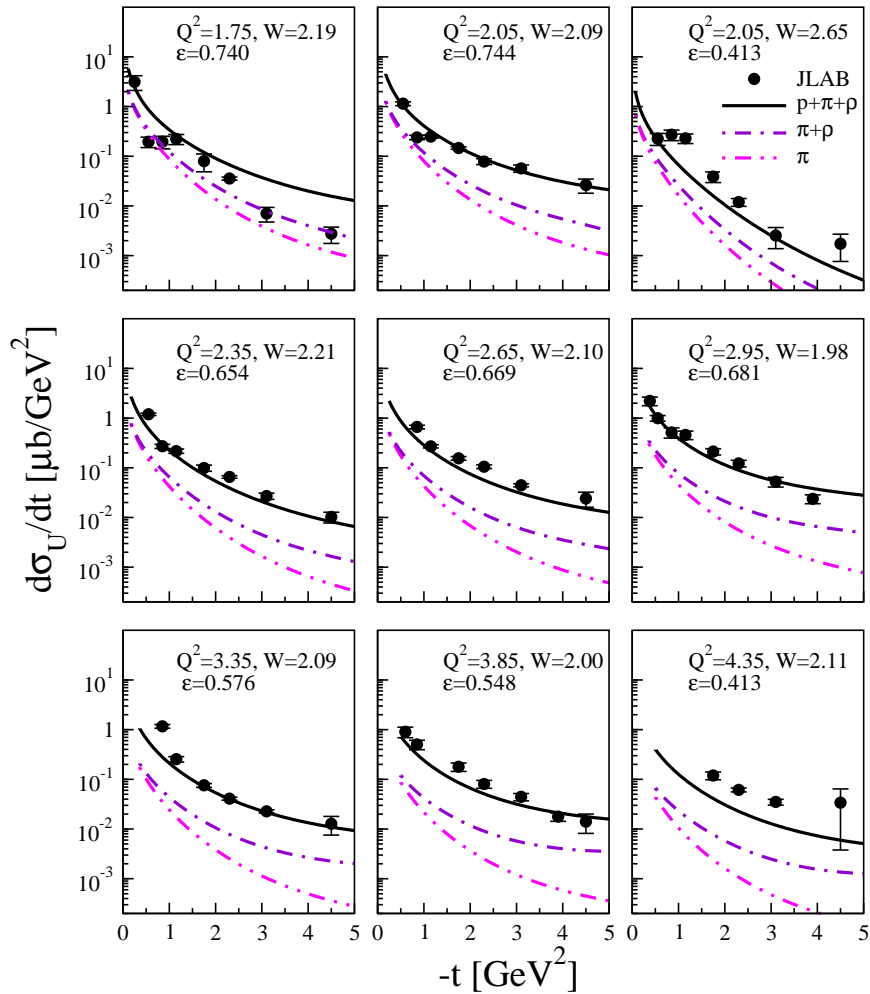}
    \caption{(Color online)
    Contribution of meson exchange in the unseparated cross section $d\sigma_U /dt$.
    Dash-dot-doted line is the contribution from $\pi$ exchange.
    Dash-dotted line corresponds to the contribution of
    $\pi+\rho$. Solid line results from $p({\rm proton})+\pi + \rho$ exchanges
    with $\Lambda_\pi=0.65$ GeV, and $\Lambda_1=1.55$ GeV.
    Data are taken from Ref. \cite{park}.} \label{clasdsdtu}
\end{figure*}

The proton exchange in the $s$-channel, though added to restore
gauge invariance, plays a role in the transverse cross section
$d\sigma_T/dt$, which corresponds to photoproduction of $\pi^+$ at
the photon point $Q^2=0$. It is reasonable to assume $N^*$
contributions in the resonance region \cite{kaskulov,vrancx}. The
$Q^2$ dependence of the proton form factor, called the Dirac form factor,
$F_1(Q^2)$
%and Pauli form factor $F_2(Q^2)$
is determined from the Sachs electric and magnetic
form factors $G_E(Q^2)$ and $G_M(Q^2)$ by using the relation
\begin{eqnarray}\label{diracff}
F_1 (Q^2)=\frac{G_E(Q^2)+\tau G_M(Q^2)}{1+\tau},
% \,\,\,\, \kappa F_2(Q^2)=\frac{G_M(Q^2)-G_E(Q^2)}{1+\tau}
\end{eqnarray}
where $\tau (Q^2)=Q^2/4M_P^2$ and $F_1(0)=1$. In the measurement
of  the on-shell form factors of a nucleon, $G_E(Q^2)$ and $G_M(Q^2)$
are applied and parameterized as
\begin{eqnarray}\label{sachff}
G_E(Q^2)=G_D(Q^2) \,, \,\,\,\, G_M(Q^2)=\mu_p G_D(Q^2),
\end{eqnarray}
where
\begin{eqnarray}\label{dpff}
G_D(Q^2)=\left(1+Q^2/\Lambda_1^2 \right)^{-2}
\end{eqnarray}
of the dipole type with $\Lambda_1^2 =0.71$ GeV$^2$
%($\Lambda_1=0.84$ GeV)
fitted to the empirical data \cite{hand}. $G_{E}(0)=1$ and
$G_{M}(0)=\mu_p=2.793$ normalized for the proton state. As
mentioned in the beginning, we adopt in this work the proton charge
form factor $F_1(Q^2)$ given in Eq. (\ref{diracff}), together with
Eqs. (\ref{sachff}), and (\ref{dpff}) above. Note that by
the definition in Eq. (\ref{diracff}), followed by Eqs. (\ref{sachff}) and
(\ref{dpff}), the proton charge form factor $F_1(Q^2)$ in this work
differs not only by the cutoff mass $\Lambda_{\gamma pp^*}(s)$
but also by the overall factor $\left(1+\tau\mu\over
1+\tau\right)$ from that of VR \cite{vrancx} given in Eq.
(\ref{vrff}).
%\begin{eqnarray}
%F_s(Q^2)=\left(1+\tau\mu\over 1+\tau\right)\left(
%1+{Q^2\over\Lambda_1^2}\right)^{-2}
%\end{eqnarray}

In Fig. \ref{emffs}, we present the $Q^2$-dependence of the proton charge
form factors $F_1(Q^2)$ adopted in the models of KM, VR, and the
present work for comparison. The dashed line describes the form factor
$F_1(Q^2)$ in Eq. (\ref{dpff}) with $\Lambda_1 = 0.84$ GeV, which
is in good agreement with experimental data of Refs.
12 and 13. %\cite{perdrisat,kelly04}.
The dotted line  is from the transition form
factor $F_s(Q^2, s\rightarrow M_p^2)$  in the KM model (Eq. (43) of
Ref 5) %\cite{kaskulov})
%with $\Lambda_1\approx 1.48$ GeV in the limit $s\rightarrow M_p^2$
while the dash-dotted line is from the form
factor in Eq. (\ref{vrff}) for the VR model. The solid line results
from the Dirac form factor $F_1(Q^2)$ with $\Lambda_1 = 1.55$ GeV used
in the present work.

%%% numerical results %%%

%Numerical analyses are presented for four differential cross
%sections in Eq. (\ref{crss1}) of the process $p(e,\,e'\pi^+)n$ in
%KM, VR, and present model for comparison.
While differences in the
proton form factor among the models are apparent, the cutoff mass
$\Lambda_\pi$ based on the pion form factor in Eq. (\ref{piff0}),
common to all models, should be different from one another for an
agreement with experiment. The KM model divides the cutoff mass into three
parts: $\Lambda_\pi=0.775$ GeV for $Q^2<0.4$ GeV$^2$, 0.63 GeV for
$0.6<Q^2<1.5$ GeV$^2$, and 0.68 GeV for $Q^2> 1.5$ GeV$^2$. The VR
model fixed $\Lambda_\pi=0.655$ GeV, regardless of the $Q^2$ range with a
smaller coupling constant $g_{\pi NN}=13.0$. We use
$\Lambda_\pi=0.78$ GeV for the DESY data and $\Lambda_\pi=0.65$ GeV
for the JLab data with the coupling constants concerning $\pi$
and $\rho$ being the same as those of the KM model.

Figure \ref{desy21} shows four differential cross sections
resulting from the KM model, the VR model, and the model used in this work for the small $-t$ and $Q^2$
region. Note that these model predictions are made within our
simple framework of ${\rm proton}+\pi+\rho$ exchanges where neglecting
the higher-spin meson exchange employed in the KM and the VR models, which is
by two orders of magnitude smaller than $\pi+\rho$. Therefore, the
differences between the models are basically the differences of the proton
form factors together with the cutoff $\Lambda_\pi$, as discussed
above. In this analysis, we fix the cutoff mass $\Lambda_1=1.55$
GeV over whole range of $Q^2$ and $-t$ for the proton form
factor in Eq. (\ref{diracff}) while the VM model uses $\Lambda_{\gamma
pp^*}(s)=0.84+(2.19-0.84)(1-M_p^2/s)$(GeV). The KM form factor
would yield $\Lambda_1\approx 1.48$ GeV in the limit of $s\to
M_p^2$, as estimated in correspondence with the same form factor
for the VM model \cite{vrancx}. Thus, all the cutoff masses for the proton form
factors in these models are almost two times larger than the
$\Lambda_1=0.84$ GeV fitted to the on-shell form factor from
experimental data, suggestive of  an $N^*$ contribution or an off-shell
effect in the $s$-channel.

\begin{figure}[]
\centering
      \includegraphics[width=0.83\columnwidth]{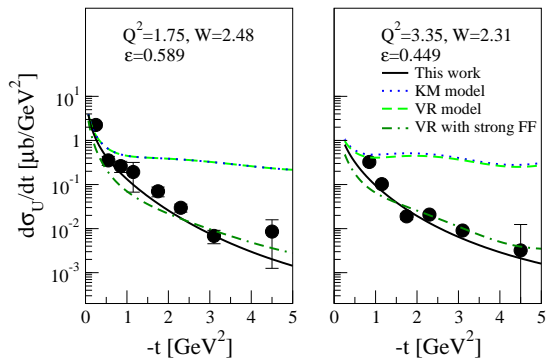}
    \caption{(Color online)  $-t$ dependence of unseparated cross section $d\sigma_U /dt$.
Notations are the same as in Fig. \ref{jlab3exp} for  solid,
dotted and  dashed lines. The dash-dash-dotted line is from the VR
model with the inclusion of the strong hadronic form factor given
in Eq. (28) of Ref.\cite{vrancx}. Data are taken from Ref.
\cite{park}. }\label{clasdsdtumc}
\end{figure}

In the high-$Q^2$ region, the JLab data for small $-t$ and large $-t$
are presented in Fig. \ref{jlab3exp} and Fig. \ref{clasdsdtu}. Our
model reproduces those cross sections with a reduced value
$\Lambda_\pi=0.65$ GeV while using an fixed $\Lambda_p=1.55$ GeV, which is
independent of $Q^2$ and $-t$ momentum transfer.

The change in $\Lambda_\pi$ due to a variation of $Q^2$ can be
understood as a change in the measurement of the charge radius of pion,
\begin{eqnarray}
<r_\pi^2>=-6 \frac{dF_\pi}{dQ^2}\bigg|_{Q^2=0} =6/\Lambda_\pi^2,
\end{eqnarray}
which reads $<r_\pi^2>=0.384$ fm$^2$ for $\Lambda_\pi=0.78$ GeV
and $<r_\pi^2>=0.55$ fm$^2$ for $\Lambda_\pi=0.65$ GeV. This is
sensible because the size of the pion will increase with
higher resolution as $Q^2$ increases.

Based on the present model with cutoff masses consistent with DESY
and JLab data, we analyze the unseparated cross section
\begin{eqnarray}
d\sigma_U=d\sigma_T+\epsilon d\sigma_L
\end{eqnarray}
at the high $Q^2$ and large $-t$ region. We note in Fig.
\ref{clasdsdtu} that the contribution of $\rho$ exchange becomes
noticeable in this kinematical region. We should point out
that the sign of $\gamma\rho\pi$ coupling is of importance in
reproducing the data. A comparison of model predictions is given  in
Fig. \ref{clasdsdtumc} for the cross section at high $Q^2$ and
$-t$ up to 5 GeV$^2$. The result confirms the present work to be
more reliable for wider applications than the others.

%--------------------------------------------------
\section*{Acknowledgment}
%--------------------------------------------------
This work was partly supported  by grant NRF-2013R1A1A2010504
from the National Research Foundation (NRF) of Korea.

\end{document}